\documentclass[amsmath,amssymb,twocolumn,showpacs]{revtex4}
\usepackage{graphicx}
\usepackage{dcolumn}

\begin{document}

\title{Leptogenesis via multiscalar coherent evolution
\\
with supersymmetric neutrino see-saw}

\author{Masato Senami}
\altaffiliation{E-mail address: senami@nucleng.kyoto-u.ac.jp}
\author{Katsuji Yamamoto}
\altaffiliation{E-mail address: yamamoto@nucleng.kyoto-u.ac.jp}
\affiliation{Department of Nuclear Engineering, Kyoto University,
Kyoto 606-8501, Japan}

\date{\today}

\begin{abstract}
A novel scenario of leptogenesis is investigated
in the supersymmetric neutrino see-saw model.
The right-handed sneutrino $ {\tilde N} $
and the $ \phi $ field in the $ {\tilde L} H_u $ direction
of the slepton and Higgs doublets
start together coherent evolution after the inflation
with right-handed neutrino mass $ M_N $
smaller than the Hubble parameter of inflation.
Then, after some period the motion of $ {\tilde N} $ and $ \phi $
is drastically changed
by the cross coupling $ M_N h_\nu {\tilde N}^* \phi \phi $
from the $ M_N N N $ and $ h_\nu N L H_u $ terms,
and the significant asymmetries of $ {\tilde N} $ and $ {\tilde L} $
are generated.
The $ {\tilde L} $ asymmetry is fixed later by the thermal effect
as the lepton number asymmetry for baryogenesis,
while the $ {\tilde N} $ asymmetry disappears through the decays
$ {\tilde N} \rightarrow {\bar L} {\bar{\tilde H}}_u , {\tilde L} H_u $
with almost the same rate but opposite final lepton numbers.
\end{abstract}
\maketitle

Baryogenesis via leptogenesis is considered
as one of the promising scenarios
to explain the baryon number asymmetry in the universe
\cite{FY}.
The leptogenesis is interesting particularly in the point
that it may be related to the neutrino mass generation.
In the supersymmetric standard model, as investigated fully so far,
the leptogenesis may be realized via the Affleck-Dine mechanism
\cite{AD,DRT}
in the $ {\tilde L} H_u $ flat direction
of the slepton and Higgs doublets, $ {\tilde L} $ and $ H_u $,
requiring the very small mass of the lightest ordinary neutrino
\cite{AFHY}.
It is also possible to realize the leptogenesis
on the flat manifold of $ {\tilde L} $-$ H_u $-$ H_d $,
where the restriction on the lightest neutrino mass
may be considerably moderated
\cite{flatmanifold}.
In this letter, we investigate another novel scenario of leptogenesis
in the supersymmetric see-saw model for neutrino masses
\cite{seesaw}.
The lepton number asymmetry is indeed generated
via the coherent evolution of the multiscalar fields,
the right-handed sneutrino $ {\tilde N} $
and the $ \phi $ field in the $ {\tilde L} H_u $ direction.
The potential terms provided with the supersymmetric neutrino see-saw
and also the thermal effect \cite{thermaleffect}
play important roles for leptogenesis in the respective epochs.
The leptogenesis is completed when the generated lepton number asymmetry
is fixed to some significant value by the thermal effect
at the scale much higher than the gravitino mass
$ m_{3/2} \sim 10^3 {\rm GeV} $.
Hence, this scenario is not restricted
by the low-energy electroweak physics.

In the present scenario of leptogenesis, it is supposed that
some of the masses $ M_N $'s of the right-handed neutrinos $ N $'s
(antineutrinos strictly) are smaller than the Hubble parameter
$ H_{\rm inf} \sim 10^{13} {\rm GeV} $ during the inflation.
Specifically, we describe the generation of lepton number asymmetry
by considering the simple case that
\begin{equation}
M_{N_1} < H_{\rm inf}
\label{MN123}
\end{equation}
for one $ N_1 $
while $ M_{N_2} , M_{N_3} > H_{\rm inf} $ for the others $ N_2 , N_3 $.
The lepton doublets are arranged with unitary transformation
so that only $ L_1 $ has the Yukawa coupling with $ N_1 $.
Then, the right-handed sneutrino $ {\tilde N}_1 $
and the $ \phi $ field in $ {\tilde L}_1 H_u $
start together coherent evolution with large initial field values
after the inflation in the manner of Affleck and Dine.
The motions of $ {\tilde N}_1 $ and $ \phi $ are linked
through the superpotential term $ h_\nu N_1 L_1 H_u $.
($ {\tilde N}_2 = {\tilde N}_3 = 0 $ due to the large masses
$ M_{N_2} , M_{N_3} > H_{\rm inf} $.)
Henceforth the generation indices are suppressed
by considering only the one generation for leptogenesis,
and the relevant scalar fields are specified as
\begin{equation}
{\tilde N} ,
{\tilde L} = \left( \begin{array}{c} \phi / {\sqrt 2} \\ 0
\end{array} \right) ,
H_u = \left( \begin{array}{c} 0 \\ \phi / {\sqrt 2} \end{array} \right) .
\label{scalars}
\end{equation}
If some of $ M_N $'s are smaller than $ H_{\rm inf} $ in general,
the coherent evolution after the inflation may be much more multidimensional
involving the $ {\tilde N} $'s, $ {\tilde L} $'s and $ H_u $.
The leptogenesis scenario is essentially valid even in such cases,
where the main source for asymmetry generation
is the cross coupling $ M_N h_\nu {\tilde N}^* \phi \phi $
from the $ M_N N N $ and $ h_\nu N L H_u $ terms
of supersymmetric neutrino see-saw.

The relevant superpotential is given by
\begin{eqnarray}
W &=& \frac{M_N}{2} N N + \frac{{\rm e}^{i \delta_N}}{4M} N N N N
+ h_\nu N L H_u .
\label{W}
\end{eqnarray}
The $ NNNN $ term may originate in the physics of Planck scale,
and its phase factor $ {\rm e}^{i \delta_N} $ is included here
with real $ M_N $.
The $ NNN $ term is discarded for simplicity by requiring the $ R $-parity.
The $ L H_u L H_u $ term is not considered either,
since it does not provide significant effect
if the Yukawa coupling $ h_\nu > 0 $ is not extremely small.
As seen later in Eqs. (\ref{N0phi0}) and (\ref{hnu}),
the Yukawa coupling $ h_\nu \sim 3 \times 10^{-3} $
is relevant for the present scenario of leptogenesis
starting at the large scale $ \sim 10^{15} {\rm GeV} $.
Then, its value at the electroweak scale $ M_W $ is evaluated
as $ h_\nu ( M_W ) \sim 10^{-3} $
by considering the renormalization group effects
mainly provided by the top quark loop for the $ H_u $ field.
The ordinary neutrino mass via see-saw mechanism is roughly estimated as
\begin{eqnarray}
m_\nu \sim 10^{-4} {\rm eV}
\left( \frac{h_\nu ( M_W )}{10^{-3}} \right)^2
\left( \frac{10^{11} {\rm GeV}}{M_N} \right)
\end{eqnarray}
depending on $ h_\nu ( M_W ) $ and $ M_N $.
(The neutrino mixing is present in general with matrix form of $ h_\nu $.)
Hence, this neutrino relevant for leptogenesis
should be identified with the lightest one,
being compatible with the data on the atmospheric
and solar neutrino experiments
\cite{SK,SNO}.
It is interesting that the lightest neutrino mass
is expected to be $ m_\nu \sim 10^{-4} {\rm eV} $
for the present leptogenesis with $ {\tilde N} $ and $ \phi $,
while $ m_\nu \lesssim 10^{-8} {\rm eV} $ is required
for the conventional Affleck-Dine leptogenesis
in the $ {\tilde L} H_u $ flat direction
\cite{AFHY}.

The scalar potential is given with $ W $ in Eq. (\ref{W}) as
\begin{eqnarray}
V &=& - c_N H^2 | {\tilde N} |^2 - c_\phi H^2 | \phi |^2
\nonumber \\
&+& \left| M_N {\tilde N}
+ \frac{{\rm e}^{i \delta_N}}{M} {\tilde N}{\tilde N}{\tilde N}
+ \frac{h_\nu}{2} \phi \phi \right|^2
+ h_\nu^2 | {\tilde N} |^2 | \phi |^2
\nonumber \\
&+& H \left( b_N \frac{M_N}{2} {\tilde N} {\tilde N}
+ a_N \frac{{\rm e}^{i \delta_N}}{4M}
{\tilde N}{\tilde N}{\tilde N}{\tilde N}
+ {\rm h.c.} \right)
\nonumber \\
&+& H \left( a_h \frac{h_\nu}{2} {\tilde N} \phi \phi
+ {\rm h.c.} \right) + V_{\rm th} ( \phi ) .
\label{V}
\end{eqnarray}
Here the soft supersymmetry breaking terms are induced
by the expansion of the universe with the Hubble parameter $ H $.
The thermal terms \cite{thermaleffect} are also included
in $ V_{\rm th} ( \phi ) $.
The $ D^2 $ terms are vanishing for the $ \phi $ field.
The evolution of the scalar fields is governed
by the equations of motion with this potential $ V $
and the redshift of $ H $.

\begin{flushleft}
{\bf (i) Inflation epoch:} $ H = H_{\rm inf} $
\end{flushleft}

The scalar fields settle into one of the minima
$ ( {\tilde N}_{0} , \phi_0 ) $ of $ V $
during the inflation with $ H = H_{\rm inf} $,
which are determined as
\begin{eqnarray}
| {\tilde N}_{0} | \sim  | \phi_0 |
\sim 3 \times 10^{15} {\rm GeV}
\left( \frac{H_{\rm inf} M}{10^{13} {\rm GeV} 10^{18} {\rm GeV}}
\right)^{1/2}
\label{N0phi0}
\end{eqnarray}
for the Yukawa coupling
\begin{eqnarray}
h_\nu \sim 3 \times 10^{-3}
\left( \frac{H_{\rm inf} / M}
{10^{13} {\rm GeV} / 10^{18} {\rm GeV}} \right)^{1/2} .
\label{hnu}
\end{eqnarray}
(We henceforth take $ H_{\rm inf} = 10^{13} {\rm GeV} $ typically.)
We consider for definiteness the case with this range of $ h_\nu $,
though it does not require a fine tuning.
If $ h_\nu < ( H_{\rm inf} / M )^{1/2} $,
$ | {\tilde N}_{0} | $ and $ | \phi_0 | $ take larger values.
If $ h_\nu > ( H_{\rm inf} / M )^{1/2} $, on the other hand,
$ \phi_0 = 0 $ may be obtained
due to the $ h_\nu^2 | {\tilde N} |^2 | \phi |^2 $ term.
The leptogenesis can be realized even in these cases
with some modifications of scenario,
which will be described elsewhere.

\begin{flushleft}
{\bf (ii) Oscillation epoch:} $ H_{\rm inf} > H > H_{\rm tr} $
\end{flushleft}

After the inflation the Hubble parameter decreases as $ H = (2/3) t^{-1} $
in the matter dominated universe,
and the multiscalar coherent evolution of $ {\tilde N} $ and $ \phi $
starts with the initial condition
$ ( {\tilde N} , \phi ) = ( {\tilde N}_0 , \phi_0 ) $
at $ t = t_0 \sim H_{\rm inf}^{-1} $, as given in Eq. (\ref{N0phi0}).
The higher order potential terms suppressed by $ M $
are soon reduced by redshift,
and the quartic couplings $ h_\nu^2 | \phi |^4 $
and $ h_\nu^2 | {\tilde N} |^2 | \phi |^2 $ dominate in this epoch
with $ h_\nu $ as given in Eq. (\ref{hnu}).
Then, driven by these quartic couplings,
the scalar fields oscillate in magnitude with scaling by redshift as
\begin{eqnarray}
| {\tilde N} | \sim  | \phi |
\sim ( H_{\rm inf} M )^{1/2} ( H / H_{\rm inf} )^{2/3} \propto H^{2/3} .
\label{Nphi}
\end{eqnarray}
The field phases, however, remain almost constant
except for the vicinities of $ {\tilde N} = 0 $ and $ \phi = 0 $,
and the significant asymmetries
of $ {\tilde N} $ and $ {\tilde L} $ do not appear in this epoch.

\begin{flushleft}
{\bf (iii) Transition epoch:}
$ H_{\rm tr} \gtrsim H > H_{\rm th} $
\end{flushleft}

Since $ {\tilde N} $ and $ \phi $ decrease with $ H $
as given in Eq. (\ref{Nphi}),
the mass term $ M_N^2 | {\tilde N} |^2 $
and $ M_N $-$ h_\nu $ cross coupling $ M_N h_\nu {\tilde N}^* \phi \phi $
become comparable to the quartic couplings $ h_\nu^2 | \phi |^4 $
and $ h_\nu^2 | {\tilde N} |^2 | \phi |^2 $
with $ | {\tilde N} | \sim | \phi | \sim M_N / h_\nu $
and the Hubble parameter
\begin{equation}
H_{\rm tr} \sim 10^{10} {\rm GeV}
\left( \frac{M_N}{10^{11} {\rm GeV}} \right)^{3/2} ,
\label{Htr}
\end{equation}
where $ h_\nu \sim ( H_{\rm inf} / M )^{1/2} $
with $ H_{\rm inf} = 10^{13} {\rm GeV} $ is taken from Eq. (\ref{hnu}).
The thermal mass term should also be considered at $ H \sim H_{\rm tr} $,
which is given by $ ( y T_{\rm p} )^2 | \phi |^2 $
with relevant coupling $ y $
under the condition $ y | \phi | < T_{\rm p} $
\cite{thermaleffect}.
The temperature $ T_{\rm p} $ of the dilute plasma
of inflaton decay products is given in terms of the reheating temperature
$ T_{\rm R} $ of the universe after the inflaton decay is completed:%
\begin{equation}
T_{\rm p} \sim (T_{\rm R}^2 H M_{\rm P})^{1/4} ,
\label{Tp}
\end{equation}
where $ M_{\rm P} = 2.4 \times 10^{18} {\rm GeV} $
is the reduced Planck mass.
The thermal mass is constrained at $ H \sim H_{\rm tr} $
as $ y T_{\rm p} < T_{\rm p}^2 / | \phi | \sim h_\nu T_{\rm p}^2 / M_N $
for $ y | \phi | < T_{\rm p} $
with $ | {\tilde N} | \sim | \phi | \sim M_N / h_\nu $.
Hence, the thermal mass term $ ( y T_{\rm p} )^2 | \phi |^2 $
is smaller than the $ M_N^2 | {\tilde N} |^2 $ 
and $ M_N h_\nu {\tilde N}^* \phi \phi $ terms
at $ H \sim H_{\rm tr} $ for the right-handed neutrino mass
\begin{equation}
M_N \gtrsim 10^{10} {\rm GeV}
\left( \frac{h_\nu}{3 \times 10^{-3}} \right)^{4/5}
\left( \frac{T_{\rm R}}{10^9 {\rm GeV}} \right)^{4/5} .
\label{MNbound}
\end{equation}

In this situation, the $ M_N^2 | {\tilde N} |^2 $
and $ M_N h_\nu {\tilde N}^* \phi \phi $ terms
as well as the $ h_\nu^2 | \phi |^4 $ term dominate
for $ H \lesssim H_{\rm tr} $,
so that the motion of $ {\tilde N} $ and $ \phi $
is changed drastically.
Specifically, the $ {\tilde N} $ field oscillates
mainly driven by the mass term $ M_N^2 | {\tilde N} |^2 $
with $ | {\tilde N} | \propto H $.
The motion of $ \phi $ follows after $ {\tilde N} $
toward the new stable configuration
with $ ( h_\nu / 2 ) \phi \phi \simeq - M_N {\tilde N} $
so as to make $ | F_N |^2 \sim | F_\phi |^2 \ll M_N^2 | {\tilde N} |^2 $
in $ V $, where $ F_N \simeq M_N {\tilde N} + ( h_\nu / 2 ) \phi \phi $
and $ F_\phi = h_\nu {\tilde N} \phi $.
Consequently, the scalar fields decrease roughly as
\begin{eqnarray}
&& | {\tilde N} | \sim ( M_N / h_\nu ) ( H / H_{\rm tr} ) \propto H ,
\label{Ntr}
\\
&& | \phi | \sim ( M_N / h_\nu ) ( H / H_{\rm tr} )^{1/2} \propto H^{1/2}
\label{phitr}
\end{eqnarray}
with $ | F_N | \sim | F_\phi | \propto H^{3/2} $.
Through this drastic change in the multiscalar coherent evolution,
the significant asymmetries of $ {\tilde N} $ and $ {\tilde L} $ appear,
which is really seen in the rate equations
\begin{eqnarray}
\frac{d}{dt} \left( \frac{n_{\tilde N}}{H^2} \right)
& \simeq & - \frac{2}{H^2} {\rm Im} [ b_N M_N H {\tilde N} {\tilde N} ]
\nonumber \\
&-& \frac{2}{H^2} {\rm Im} \left [ M_N {\tilde N} F_N^*
+ \frac{a_h h_\nu}{2} H {\tilde N} \phi \phi \right] ,
\label{dnNdt}
\\
\frac{d}{dt} \left( \frac{n_{\tilde L}}{H^2} \right)
& \simeq & - \frac{2}{H^2} {\rm Im} \left[
\frac{h_\nu}{2} \phi \phi F_N^*
+ \frac{a_h h_\nu}{2} H {\tilde N} \phi \phi \right]
\label{dnLdt}
\end{eqnarray}
with $ n_{\tilde L} = n_{H_u} = n_\phi / 2 $.
The main sources are scaled as
$ {\rm Im} [ ( h_\nu / 2 ) \phi \phi F_N^* ] / H^2 $
$ \simeq - {\rm Im} [ M_N {\tilde N} F_N^* ] / H^2 $
$ \propto H^{5/2} / H^2 $ with $ | F_N | \propto H^{3/2} $,
and hence the asymmetries $ n_{\tilde N} $ and $ n_{\tilde L} $ oscillate
rapidly by the exchange $ {\tilde N} \leftrightarrow {\tilde L} $.
The sum $ n_{\tilde N} + n_{\tilde L} $, however,
varies rather moderately with the remaining sources $ \propto H^3 $,
since the main sources are cancelled as $ {\rm Im} [ F_N F_N^* ] = 0 $
with $ F_N \simeq M_N {\tilde N} + ( h_\nu / 2 ) \phi \phi $.

\begin{flushleft}
{\bf (iv) Completion epoch:}
$ H_{\rm th} \gtrsim H \gg m_{3/2} $
\end{flushleft}

After the transition epoch continues for some period,
the thermal log term \cite{thermaleffect} eventually becomes significant
on the evolution of $ \phi $.
It is mainly provided as
\begin{equation}
a_{\rm th} \alpha_s^2 T_{\rm p}^4
\ln ( | \phi |^2 / T_{\rm p}^2 )
\end{equation}
($ a_{\rm th} = 9/8 $)
through the modification of $ {\rm SU(3)}_C $ coupling
due to the decoupling of top quark from the plasma
with large mass $ h_t | \phi | / {\sqrt 2} > T_{\rm p} $.
This thermal log term acts as the effective mass term
for the $ \phi $ field giving
$ ( a_{\rm th} \alpha_s^2 T_{\rm p}^4 / | \phi |^2 ) \phi $
$ \propto H^{1/2} $ in $ \partial V / \partial \phi^* $.
It dominates over the term $ F_N h_\nu \phi^* $
$ \propto H^2 $ in $ \partial V / \partial \phi^* $
($ | F_N | \sim | F_\phi | = h_\nu | {\tilde N} | | \phi | $)
with the Hubble parameter
\begin{eqnarray}
H_{\rm th} & \sim & 10^7 {\rm GeV}
\left( \frac{h_\nu}{3 \times 10^{-3}} \right)^{4/3}
\nonumber \\
& \times &
\left( \frac{M_N}{10^{11} {\rm GeV}} \right)^{-1/6}
\left( \frac{T_{\rm R}}{10^9 {\rm GeV}} \right)^{4/3} ,
\label{Hth}
\end{eqnarray}
where Eqs. ({\ref{Tp}), (\ref{phitr}) and (\ref{Ntr}) are considered.
Then, the rotation of the $ \phi $ field phase is accelerated
by this thermal log term with the change of field scaling
\begin{equation}
| \phi | \propto H^{1/2} \rightarrow H^{3/2}
\end{equation}
while keeping $ | {\tilde N} | \propto H $.
After a while the top quark enters the plasma
at $ H \sim 0.1 H_{\rm th} $ with $ | \phi | \sim T_{\rm p} $
($ h_t \sim 1 $).
Then, the thermal mass term $ T_{\rm p}^2 | \phi |^2 $
instead becomes important,
and the $ \phi $ field decreases as $ | \phi | \propto H^{7/8} $.
In the preceding epoch,
the significant exchange of asymmetries,
$ n_{\tilde N} \leftrightarrow n_{\tilde L} $, took place
through the $ {\tilde N} $-$ \phi $ couplings,
as seen in Eqs. (\ref{dnNdt}) and (\ref{dnLdt}).
These couplings are actually turned off in this epoch
with the rapid decrease of $ | \phi | \propto H^{3/2} $
and $ H^{7/8} $ later due to the thermal terms,
and the $ \phi $ and $ {\tilde N} $ evolve almost independently.

In this way, the $ {\tilde L} $ asymmetry is fixed
to some significant value as the lepton number asymmetry
for $ t > H_{\rm th}^{-1} $,
\begin{equation}
n_{\tilde L} \simeq n_L
\equiv \epsilon_L [(3/2) H^2 M] ,
\end{equation}
since the $ {\tilde L} $ violating sources in Eq. (\ref{dnLdt})
decreases fast enough as $ H^4 / H^2 $ and $ H^{11/4} / H^2 $ later
with rapidly varying phase of $ {\tilde N}^* \phi \phi $.
This concludes that the thermal effect plays
the positive role for the completion of leptogenesis,
which is in salient contrast to the conventional Affleck-Dine mechanism
where the thermal effect rather suppresses the asymmetry seriously.
The resultant lepton-to-entropy ratio after the reheating
is estimated with $ s \simeq 3 H_{\rm R}^2 M_{\rm P}^2 / T_{\rm R} $ as
\begin{equation}
\frac{n_L}{s} \sim 10^{-10}
\left( \frac{\epsilon_L}{1} \right)
\left(\frac{M}{10^{18} \rm GeV }  \right)
\left(\frac{T_{\rm R}}{10^9 \rm GeV}  \right) .
\end{equation}
Here the reheating temperature is restricted
as $ T_{\rm R} \lesssim 10^8 - 10^{10} {\rm GeV} $
to avoid the gravitino problem
\cite{gravitino,gravitino2,gravitino3}.
The lepton number asymmetry is converted to the baryon number asymmetry
through the electroweak anomalous effect
as $ n_B / s = - (8 / 23) n_L / s $ \cite{harveyturner}.
Hence, the sufficient baryon-to-entropy ratio
can be provided for the nucleosynthesis
with $ \eta = (2.6 - 6.2) \times 10^{-10} $
\cite{PDG}.

The motion of $ {\tilde N} $ after the decoupling from $ \phi $
for $ H < H_{\rm th} \ll M_N $ is determined
by the $ M_N^2 | {\tilde N} |^2 $
and $ b_N H M_N {\tilde N} {\tilde N} $ terms,
and the analytic solution is obtained in a good approximation
with $ H \ll M_N $ for the two eigenmodes
$ \eta_{\rm R} (t) $ and $ \eta_{\rm I} (t) $ in $ {\tilde N} (t) $ as
\begin{equation}
\eta_{\rm R,I} (t) \simeq {\bar \eta}_{\rm R,I}
\cos \left[ M_N t + \sigma_{\rm R,I} ( | b_N | / 3 ) \ln t
+ \delta_{\rm R,I} \right]
\label{etaRI}
\end{equation}
with $ \sigma_{\rm R} = + 1 $, $ \sigma_{\rm I} = - 1 $, and
\begin{equation}
{\tilde N} \equiv H ( M / M_N )^{1/2}
( b_N / | b_N | )^{-1/2} ( \eta_{\rm R} + i \eta_{\rm I} ) .
\label{Nt}
\end{equation}
The parameters $ {\bar \eta}_{\rm R,I} $ and $ \delta_{\rm R,I} $
are determined as the result of $ {\tilde N} $ motion
from $ t = t_0 \sim H_{\rm inf}^{-1} $
through $ t > H_{\rm th}^{-1} \gg M_N^{-1} $.
The $ {\tilde N} $ asymmetry is evaluated
with Eqs. (\ref{etaRI}) and (\ref{Nt}) as
\begin{eqnarray}
n_{\tilde N} (t) \simeq - 2 H^2 M
{\bar \eta}_{\rm R} {\bar \eta}_{\rm I}
\cos [ ( 2 | b_N | / 3 ) \ln t + \delta_{\rm R} - \delta_{\rm I} ] ,
&&
\label{nN}
\end{eqnarray}
where the rapid oscillations of $ \eta_{\rm R} $ and $ \eta_{\rm I} $
with $ M_N t $ in Eq. (\ref{etaRI}) are cancelled.

This $ {\tilde N} $ asymmetry oscillates slowly in $ \ln t $
for some while due to the $ b_N $ term, as seen in Eq. (\ref{nN}).
Then, the incoherent decays of $ {\tilde N} $ become significant
with the dominant modes
\begin{equation}
{\tilde N} \rightarrow {\bar L} {\bar {\tilde H}_u} [ L = - 1 ] , \
 {\tilde L} H_u [ L = + 1 ] ,
\label{Ndecays}
\end{equation}
where the decay products are ultra-relativistic
with $ h_\nu | {\tilde N} | \ll M_N / 2 $.
The motion of $ {\tilde N} $ is significantly
decelerated by these $ {\tilde N} $ decays
at $ H \sim \Gamma_{\tilde N} \simeq ( h_\nu^2 / 4 \pi ) M_N $
($ \sim 10^5 {\rm GeV} $ numerically),
so that it is linked again to $ \phi $,
tracking the instantaneous minimum of $ V $
as $ {\tilde N} \simeq - ( h_\nu / 2 M_N ) \phi \phi $
with $ | {\tilde N} | \propto | \phi |^2 \propto H^{7/4} $ in magnitude
and $ d \theta_{\tilde N} / dt \simeq 2 d \theta_\phi / dt $
$ \propto H^{1/4} $ in phase.
Then, the $ {\tilde N} $ asymmetry remaining after the transition epoch
diminishes rapidly through the decays
as $ n_{\tilde N} = 2 ( d \theta_{\tilde N} / dt ) | {\tilde N} |^2 $
$ \propto H^{15/4} $.
It is the essential point that the decay modes (\ref{Ndecays})
have almost the same rate $ \Gamma_{\tilde N} / 2 $
but the opposite final lepton numbers $ L = \pm 1 $.
This means that the $ {\tilde N} $ asymmetry
does not leave any significant lepton number asymmetry.

\begin{figure}[t]
\begin{center}
\scalebox{.55}{\includegraphics*[0cm,0cm][18cm,10cm]{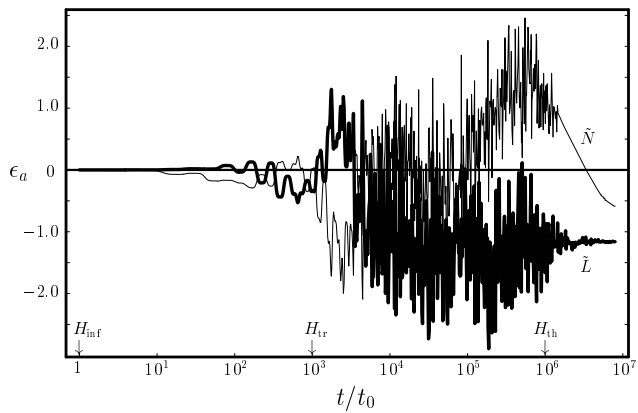}}
\caption{Typical time variations of the asymmetries
$ \epsilon_{\tilde N} (t) $ (thin)
and $ \epsilon_{\tilde L} (t) \simeq \epsilon_L (t) $ (bold)
are depicted.}
\label{asymm}
\end{center}
\end{figure}

The equations of motion for $ {\tilde N} $ and $ \phi $
are solved by numerical calculations
to confirm the present scenario of leptogenesis.
The typical time variations of $ n_{\tilde N} (t) $
and $ n_{\tilde L} (t) \simeq n_L (t) $ are depicted in Fig. \ref{asymm}
in terms of the asymmetry fractions
$ \epsilon_a \equiv n_a / [(3/2) H^2 M] $.
Here the model parameters are taken for example as
$ H_{\rm inf} = 10^{13} {\rm GeV} $,
$ M = 5 \times 10^{18} {\rm GeV} $, $ M_N = 10^{11} {\rm GeV} $,
$ {\rm e}^{i \delta_N} = {\rm e}^{i (3/10) \pi} $,
$ h_\nu = 3 \times 10^{-3} $,
$ c_N = 1.2 $, $ c_\phi = 0.8 $,
$ b_N = 1.3 {\rm e}^{i (2/3) \pi} $,
$ a_N = 1.5 {\rm e}^{i (5/4) \pi} $,
$ a_h = 0.8 {\rm e}^{i (1/4) \pi} $,
$ T_{\rm R} = 10^9 {\rm GeV} $.
The relevant scales, $ H_{\rm inf} $, $ H_{\rm tr} $ and $ H_{\rm th} $,
are marked together specifying the respective epochs.
We really observe the expected changes of the asymmetries
through $ H_{\rm inf} \rightarrow H_{\rm tr} \rightarrow H_{\rm th} $,
resulting in the desired lepton number asymmetry $ \epsilon_L \sim 1 $.
Particularly, the variations of $ \epsilon_{\tilde L} $
and $ \epsilon_{\tilde N} $ are separated for $ t > H_{\rm th}^{-1} $;
$ \epsilon_{\tilde L} $ is fixed to some significant value
while $ \epsilon_{\tilde N} $ oscillates slowly in $ \ln t $
as given in Eq. (\ref{nN}).
It is also checked that the sum $ n_{\tilde N} + n_{\tilde L} $
varies rather moderately in the transition epoch,
while the respective asymmetries oscillate rapidly.

In summary, we have investigated the leptogenesis via multiscalar
coherent evolution in the supersymmetric see-saw model.
The right-handed sneutrino $ {\tilde N} $
and the $ \phi $ field in $ {\tilde L} H_u $
of the slepton and Higgs doublets start together coherent evolution 
after the inflation with $ M_N $ smaller than $ H_{\rm inf} $.
Then, after some period the motion of $ {\tilde N} $ and $ \phi $
is drastically changed
by the cross coupling $ M_N h_\nu {\tilde N}^* \phi \phi $,
and the significant asymmetries of $ {\tilde N} $ and $ {\tilde L} $
are generated.
The $ {\tilde L} $ asymmetry is fixed later by the thermal effect
as the lepton number asymmetry $ n_L $.
The $ {\tilde N} $ asymmetry, on the other hand,
disappears through the incoherent decays
$ {\tilde N} \rightarrow {\bar L} {\bar{\tilde H}}_u , {\tilde L} H_u $
with almost the same rate but opposite final lepton numbers.
The sufficient amount of $ n_L $ for baryogenesis
can be obtained with the lightest neutrino mass
$ m_\nu \lesssim 10^{-3} {\rm eV} $ given by the see-saw mechanism
with the right-handed neutrino mass
$ M_N \sim 10^{10} - 10^{14} {\rm GeV} \lesssim H_{\rm inf} $.

\begin{acknowledgments}
This work is supported in part by
Grant-in-Aid for Scientific Research on Priority Areas B (No. 13135214)
from the Ministry of Education, Culture, Sports, Science and Technology,
Japan.
\end{acknowledgments}

\end{document}